\documentstyle[11pt,newpasp,twoside,epsfig]{article}

\def\edcomment#1{\iffalse\marginpar{\raggedright\sl#1\/}\else\relax\fi}
\reversemarginpar

\begin{document}

\title{Mass-Loss and Magnetospheres:
X-rays from Hot Stars and Young Stellar Objects}
\author{Marc Gagn\'e}
\affil{Geology and Astronomy, West Chester University, West Chester, PA 19383}

\author{David Cohen}
\affil{Physics and Astronomy, Swarthmore College, Swarthmore, PA 19081}

\author{Stanley Owocki and Asif Ud-Doula}
\affil{Bartol Research Institute, University of Delaware, Newark, DE 19716}

\begin{abstract}
High-resolution X-ray spectra of high-mass stars and low-mass T-Tauri stars
obtained during the first year of the {\it Chandra} mission are providing
important clues about the mechanisms which produce X-rays on very young stars.
For $\zeta$~Pup (O4~If) and $\zeta$~Ori (O9.5~I), the broad, blue-shifted
line profiles, line ratios, and derived temperature distribution suggest
that the X-rays are produced throughout the wind via instability-driven
wind shocks.  For some less luminous OB stars, like $\theta^1$~Ori~C (O7~V)
and $\tau$~Sco (B0~V), the line profiles are symmetric and narrower.
The presence of time-variable
emission and very high-temperature lines in $\theta^1$~Ori~C and $\tau$~Sco
suggest that magnetically confined wind shocks may be at work.
The grating spectrum of the classical T-Tauri star TW~Hya is remarkable
because the forbidden-line emission of He-like Ne~IX and O~VII is very weak,
implying that the X-ray emitting region is very dense,
$n_{\rm e} \approx 6\times 10^{12}{\rm~cm}^{-3}$,
or that the X-rays are produced
very close to the ultraviolet hotspot at the base of an accretion funnel.
ACIS light curves and spectra of flares and low-mass and high-mass
young stellar objects in Orion and $\rho$~Ophiuchus further suggest that
extreme magnetic activity is a general property of many very young stars.
\end{abstract}
\keywords{stars: individual: $\zeta$~Puppis, $\zeta$~Orionis, $\theta^1$~Orionis, $\tau$~Scorpii, TW~Hydrae, [GY92] 195 -- X-rays: stars -- stars: pre--main-sequence -- stars: early-type}

\section{Introduction}

One of the early surprises from the {\it Einstein} mission was the detection of
strong X-ray emission from O stars (Harnden et al. 1979).
While the wind-shock paradigm for X-ray emission from hot stars (Lucy and
White 1980; Owocki, Castor, \& Rybicki 1988; MacFarlane \& Cassinelli 1989;
Hillier et al. 1993; Feldmeier et al. 1995; Cohen et al. 1996;
Owocki \& Cohen 1999) has gained
acceptance over the past two decades, there is increasing evidence that not all
early-type stars have X-ray properties that can be explained by instability-
generated wind shocks alone.  This evidence includes very hot plasma on the
young B0~V star $\tau$~Sco (Cohen, Cassinelli, \& Waldron 1997), overall
levels of X-ray emission that approach or even exceed reasonable estimates
of the wind emission measure in early- and mid-B stars (Cohen, Cassinelli, \&
MacFarlane 1997), and strong 15.4-day periodic X-ray emission on the
central O star of the Orion Nebula, $\theta^1$~Ori~C (Gagn\'e et al. 1997).

\begin{figure}
\plotone{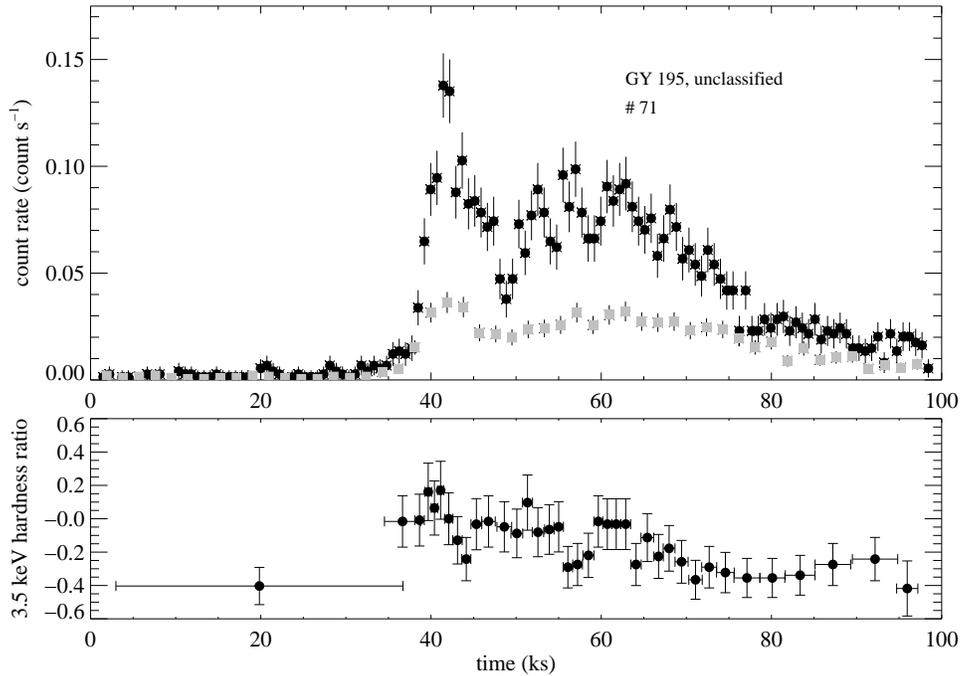}
\caption{Hard-band (2.5-8.0~keV: black circles) and soft-band
(0.5-2.5~keV gray squares) light curves of a flare on the T Tauri star (GY 195)
in the $\rho$~Oph A cloud.  The ACIS flare spectrum can be fit
with a single-temperature $4.7\pm0.4{\rm~keV}$, low-Fe abundance
($Z=0.25$) VMEKAL plasma with $N_{\rm H} = 3.9\pm0.1\times10^{22}{\rm~cm}^{-2}$.
The hard-band emission rises quickly at the onset of the flare and again
during subsequent reheating events seen in the hardness ratio time-series
(lower panel).
}
\end{figure}

The lack of outer convection zones in hot stars, and the lack of any
correlation between X-ray activity and rotation (Pallavicini et al. 1981)
has been taken to mean that the X-rays from early-type stars are not produced
by magnetic activity associated with a solar-type dynamo.
However, it has been suggested that large-scale magnetic fields, in conjunction
with radiation-driven winds, could play a role in the X-ray production on
some of these early-type stars (e.g., Babel \& Montmerle 1997a).  
The strong and hard X-ray emission from
$\theta^1$~Ori~C, $\tau$~Sco, as well as several other young hot stars
(Corocoran et al. 1994; Yamauchi et al. et al. 1996; Schulz et al. 2000),
suggest that some type of magnetic
confinement, possibly associated with youth, may lead
to the efficient production of hard X-rays on some hot stars.
If the magnetic fields on hot stars survive the star-formation process,
the transfer of angular momentum via a magnetized wind would lead to
significant spin down. 

The discovery of strong variable X-ray emission from classical T-Tauri
stars (CTTS) in Taurus-Auriga with {\it Einstein} (Gahm et al. 1980) was
less surprising because TTSs were known to be relatively fast rotators,
strong H$\alpha$ sources, and to undergo strong white-light flares.
Since then, large-amplitude, long-duration flares have been observed many 
times with {\it Einstein}, {\it ROSAT}, {\it ASCA}, and {\it Chandra}.
The temperature of the flaring plasma rises from
a quiescent value of 6--20~MK to a peak value of 50--100~MK in minutes to hours
and decays on time scales of many hours.  
The central question is whether TTSs have very active solar-type coronae
driven by a magnetic dynamo or whether flares are produced in some kind of
a magnetosphere coupled to a disk. 

The X-ray flare spectra and light curves have been modeled in two ways:
as a single heating event in a quasi-statically cooling magnetic loop and
as a series of heating events in a coronal active region.
E.g., a giant {\it ASCA} flare on the weak-line T-Tauri star (WTTS) V~773 Tau
showed a peak temperature of 100~MK and a decay time of 2.3 hours 
(Tsuboi et al. 1998).  A quasi-static cooling loop approximation
implies moderate electron densities like those observed in solar flares and
very large loop lengths, $l \ga R_\star$. Favata, Micela, \& Reale (2001)
have modeled the same flare data hydrodynamically to show that such
flares may be produced in much smaller loops. In this model, long-duration
flares are a manifestation of continuous reheating events rather than a
single flare in a large, low-density loop.

{\it ASCA} flares on three young stellar objects (YSOs) in $\rho$~Ophiuchus,
YLW~15, WL~6, and Elias~29, showed evidence of rotational modulation,
quasi-periodic flaring, and significant spectral time variability
(Tsuboi et al. 2000). Montmerle et al. (2000) present a magnetosphere/disk
interaction model to explain the quasi-periodic flares on the
Class I YSO YLW 15. Again, Favata, et al. (2001) show that such flares
may reflect reheating events in high-latitude active regions.

%In many respects, the WTTSs have magnetic activity like RS~CVn binaries.  
%WTTSs discovered in the X-ray, optical, and near-infrared now outnumber
%classical T Tauri stars (CTTS) with accretion disks by at least 2:1
%(e.g., Wichmann et al. 1996).
%As a group, WTTSs are more X-ray luminous than their CTTS counterparts.
%This difference may be caused by increased circumstellar absorption or it
%may signal a fundamentally different X-ray emission mechanism. 
%For all classes of YSOs, X-ray luminosity is fairly well
%correlated with bolometric luminosity, 
%$10^{-4} \la L_{\rm X}/L_{\rm bol} \la 10^{-3}$,
%with many WTTSs close to the magnetic activity saturation limit of $10^{-3}$.
%Plots of X-ray luminosity or X-ray surface flux versus rotation rate
%or rotational velocity generally do not show the tight activity-rotation
%correlation seen in older late-type main-sequence stars (e.g., Gagn\'e et al. 1995),
%although a sample of young ($\la 2$~Myr) TTSs in Taurus-Auriga
%shows some correlation (Bouvier 1990; Damiani \& Micela 1995).

The recent discovery of X-ray emission from deeply embedded submillimeter/radio
dust condensations in the Orion molecular cloud (Tsuboi et al. 2001) indicates
that magnetic heating begins at the earliest stages of star formation.  
It remains to be seen whether the large-scale magnetic fields which are
believed to regulate accretion, mass-loss, and bipolar outflows are
related to the reconnection events which lead to flares and X-ray activity.

Disk-photosphere coupling via a magnetic field was first suggested
by K\"onigl (1991) (inspired by the model of Ghosh \& Lamb 1979) to
explain the slow rotation of CTTSs.  A recent study of low-mass 1-30~Myr
stars in the Orion Nebula Cluster and NGC 2264 (Rebull et al. 2001) found
that PMS stars evolve with nearly constant angular velocity as they contract
down their convective tracks.  This means that the mechanism
which regulates the spin of a CTTS may still be at work on WTTSs
long after a dusty disk has been dissipated.  If magnetic coupling
still operates long into the WTTS phase, then a gas (plasma) disk must exist
within a few stellar radii of the photosphere.

In this paper we will discuss recent {\it Chandra} results which address
magnetic activity on young high-mass and low-mass stars.
\S2 summarizes results based on {\it ASCA} and {\it Chandra} light curves and
CCD spectra of YSO flares in $\rho$~Ophiuchus.  In \S3 we describe
the X-ray grating spectrum of the CTTS TW Hya and use the
forbidden lines of Ne~IX and O~VII to suggest a location for the X-ray
emitting plasma.  In \S4 we describe grating spectra of four hot
stars that, in some cases, provide strong evidence for magnetic confinement.
In \S5 we present 2-D MHD simulations of magnetically confined wind shocks.
Finally, we conclude that magnetic fields, disks, and mass-loss may play a
central role in the production of X-rays on very young high-mass and low-mass
stars.

\section{Young Stellar Objects in $\rho$~Ophiuchus}

The $\rho$~Ophiuchus A cloud was observed continuously for 96~ks with ACIS-I
on 2000 May 15.  An RGB image of the fiterered event data with 
red=soft (0.5-1.5~keV), green=medium (1.5-2.5~keV), and blue=hard (2.5-8~keV)
shows 76 X-ray sources. Highly absorbed YSOs and background AGN seen
through 25-75 mag of visual absorption appear as blue sources in this
image.  Twenty-two background AGN have been identified as faint sources
with high 2-keV hardness ratios and low Kolmogorov-Smirnov variability
statistics.  By dividing the event data into 31 time bins and removing 
flickering background pixels, we have produced a color movie illustrating
what Montmerle et al. (1983) described as an ``X-ray Christmas Tree'' 
in $\rho$~Oph.

In Figure~1, we present ACIS medium- and hard-band light curves
of the low-mass YSO GY~195 in $\rho$~Oph~A.  
The lower panel is the time-series of 3-keV hardness ratios for GY~195.
This flare is interesting because of the deep dip in the hard-band
($3-8$~keV) light curve approximately $48$~ks after the start of the
observation. Initially, we interpreted this as evidence of self-eclipsing.
However, the soft-band ($0.5-3$~keV) light curve shows a much smaller dip
than expected.  Since the emission-measure distribution at the peak of a flare
is dominated by very hot plasma ($T > 50$~MK), those times of the
when the hard emission rises sharply probably correspond to
reheating events.  We see a few small flare events
before the major peak at 41~ks, a rapid decay, and a series
of reheating events at approximately 52, 62, 83, and 93~ks.
We have yet to perform a detailed spectral time-series analysis on these data,
but our initial XSPEC analysis suggests a very hot, low-abundance coronal
plasma, and high column density consistent with other embedded T Tauri stars
in $\rho$~Oph.

The $\rho$~Ophiuchus cloud contains two optically visible stars: the young,
magnetic B3 star S1~Oph and the WTTS DoAr~21. During the {\it Chandra}
observation, DoAr~21 was in the decay phase of a very large flare.
Although no major flares were seen on S1, the light curve exhibits
low-amplitude, long-term and short-term variability.  The ACIS spectrum of this
high-mass YSO is reminiscent of low-mass YSOs:
$N_{\rm H} = 4.1\pm0.5\times10^{22}{\rm~cm}^{-2}$ and $kT = 3.1\pm0.5$~keV.  
Simultaneous VLA observations at 2 and 6~cm show variable,
non-thermal radio emission, confirming its magnetic origin, first reported by
Andr\'e et al. (1993).  S1's radio and X-ray properties make it a good candidate
for magnetically confined wind shocks (see \S5).

\section{The Classical T Tauri Star TW Hya}

With the launch of {\it Chandra} and {\it Newton}-XMM, high-resolution grating
spectroscopy has become an important new tool for studying X-ray activity on
young stars. The He-like resonance (r), intercombination (i), and forbidden (f)
lines of coronal sources like AB Dor (dotted line in Fig.~2) are useful density
diagnostics because the collisional excitation rate from the metastable
$^1{\rm S}_1$ state depends on density (see Figure~1 of Gabriel \& Jordan 1969).
The critical density is reached when the collisional excitation rate equals
the forbidden $^3{\rm S}_1 \rightarrow {^1{\rm S}_0}$ photodecay rate.
For AB Dor, the forbidden-line emission is strong; the S, Si, Mg, and Ne $f/i$
ratios are close to their low-density limits.  The O~VII $f/i$ ratio yields
$\log n_{\rm e} = 10.68\pm0.33{\rm~cm}^{-3}$.

For TW HYa, a classical T Tauri star, the $f$ lines of Ne~IX (seen in Fig.~1)
and O~VII are remarkably suppressed. Kastner et al. (2001) measure
$f/i = 0.44\pm0.13$ for Ne~IX indicating
$\log n_{\rm e} = 12.75\pm0.15~{\rm cm}^{-3}$.
This is the highest coronal density measured so far using $f/i$ ratios,
exceeding that of AB Dor by two orders of magnitude, possibly
signaling a fundamental difference between coronal stars and T Tauri stars.
Kastner et al. go further, suggesting that the bulk of the X-ray emission
on TW Hya is generated at the base of the accretion funnel which couples the
accretion disk to the stellar photosphere.

\begin{figure}
\plotone{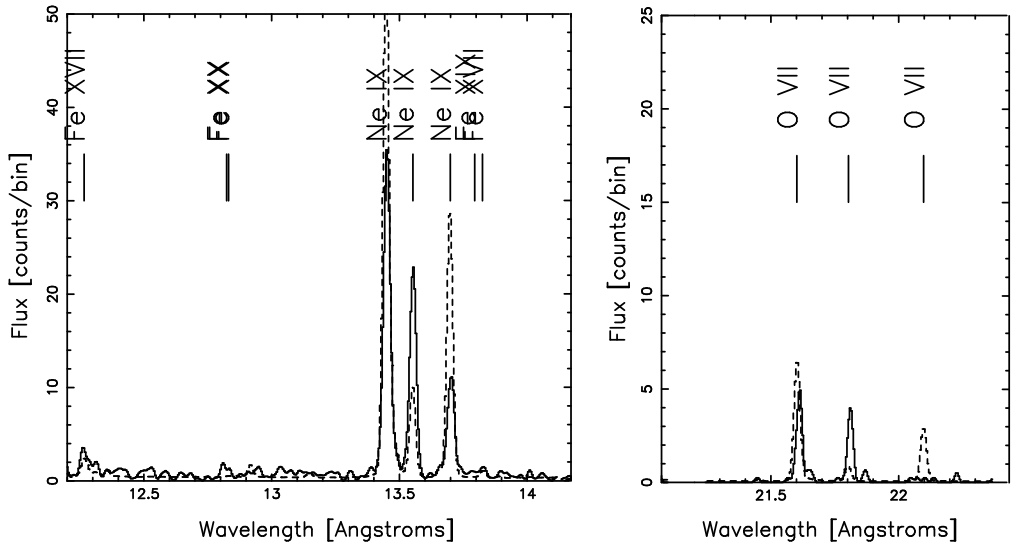}
\caption{The MEG spectrum (solid lines) of TW Hya around the $rif$
Ne~IX and O~VII triplets (Kastner et al. 2001).  The observed spectrum is
overlaid with a variable-abundance, differential emission-measure model
(dashed curve) that best-fits temperature-sensitive lines.  Note the discrepant
intercombination and forbidden lines.
}
\end{figure}

We point out, however, that ultraviolet emission can depopulate the metastable
$^3{\rm S}_1$ state via photoexcitation to the $^3{\rm P}_{210}$ state
(see Kahn et al. 2001). The forbidden line is suppressed when the
$^3{\rm S}_1 \rightarrow {^3{\rm P}}$ photoexcitation rate $R_{\rm PE}$
approaches the $^3{\rm S}_1 \rightarrow {^1{\rm S}_0}$ photodecay rate
$R_{\rm RD}$.  For O~VII and Ne~IX photoexcitation occurs in the ultraviolet at
1616 and 1246\AA.
The photoexcitation rate for the metastable state of Ne IX is
$R_{\rm PE} = F_\nu\frac{\pi e^2}{mc} f$,
where $f= 0.0949$ (Cann \& Thakkar 1992).
$R_{\rm PE} = R_{\rm RD} = 1.09\times10^4 {\rm~s}^{-1}$ (Drake 1971)
implies a critical flux
$F_\nu = 4.33\times10^{6}{\rm~s}^{-1}{\rm~cm}^{-2}{\rm~Hz}^{-1}$.
TW Hya's flux at 1245.8~\AA\ from the IUE archive is
$f_\lambda\approx3.5\times10^{-14}{\rm~ergs~s}^{-1}{\rm~cm}^{-2}{\rm~\AA}^{-1}$.
We consider the simplest geometrical model: a star at a distance $D$
(56.4~pc for TW Hya) with X-ray and UV sources separated by a distance $d$.
In this case, $F_{\nu} = ({D}/{d})^2 f_\nu$.
Solving for the UV/X-ray separation at which $R_{\rm PE} = R_{\rm PD}$,
we find $d \approx 2.5\times10^9 {\rm~cm}$, i.e., $d\approx 0.04 R_\odot$.

On TW Hya, the O~VII forbidden line is strongly suppressed.
Assuming $R_{\rm PD} = 1.04\times10^3 {\rm~s}^{-1}$ and $f = 0.0742$
implies a critical incident flux
$F_\nu = 5.28\times10^{5}{\rm~s}^{-1}{\rm~cm}^{-2}{\rm~Hz}^{-1}$.
TW Hya's IUE flux at 1616.5~\AA\ is 
$f_\lambda\approx3.0\times10^{-14}{\rm~ergs~s}^{-1}{\rm~cm}^{-2}{\rm~\AA}^{-1}$,
implying a critical UV/X-ray separation $d\approx1.1\times10^{10} {\rm~cm}$.
Since the O~VII $f$ line is entirely suppressed, $d < 0.16 R_\odot$.
Thus, the Ne~IX and O~VII
$f/i$ ratios may be suppressed via ultraviolet photoexcitation, provided
the X-rays are produced close to the UV continuum source.

\begin{figure}
\plottwo{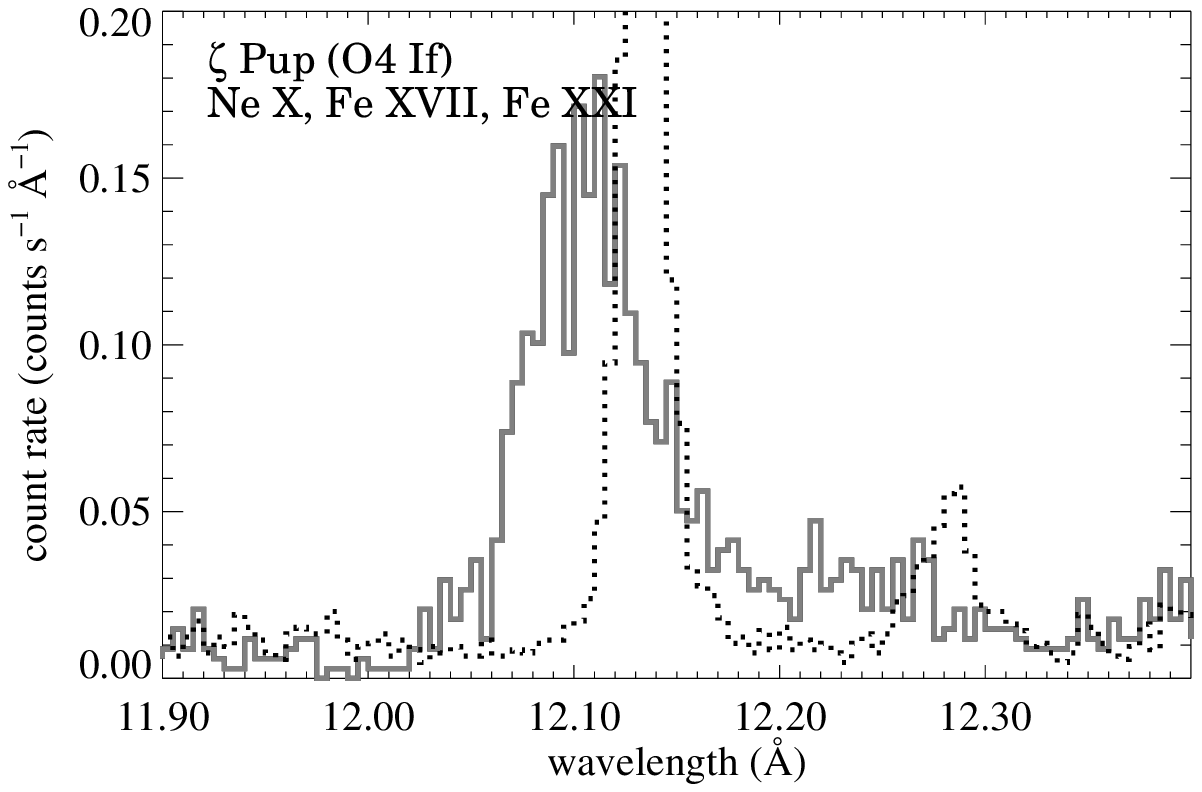}{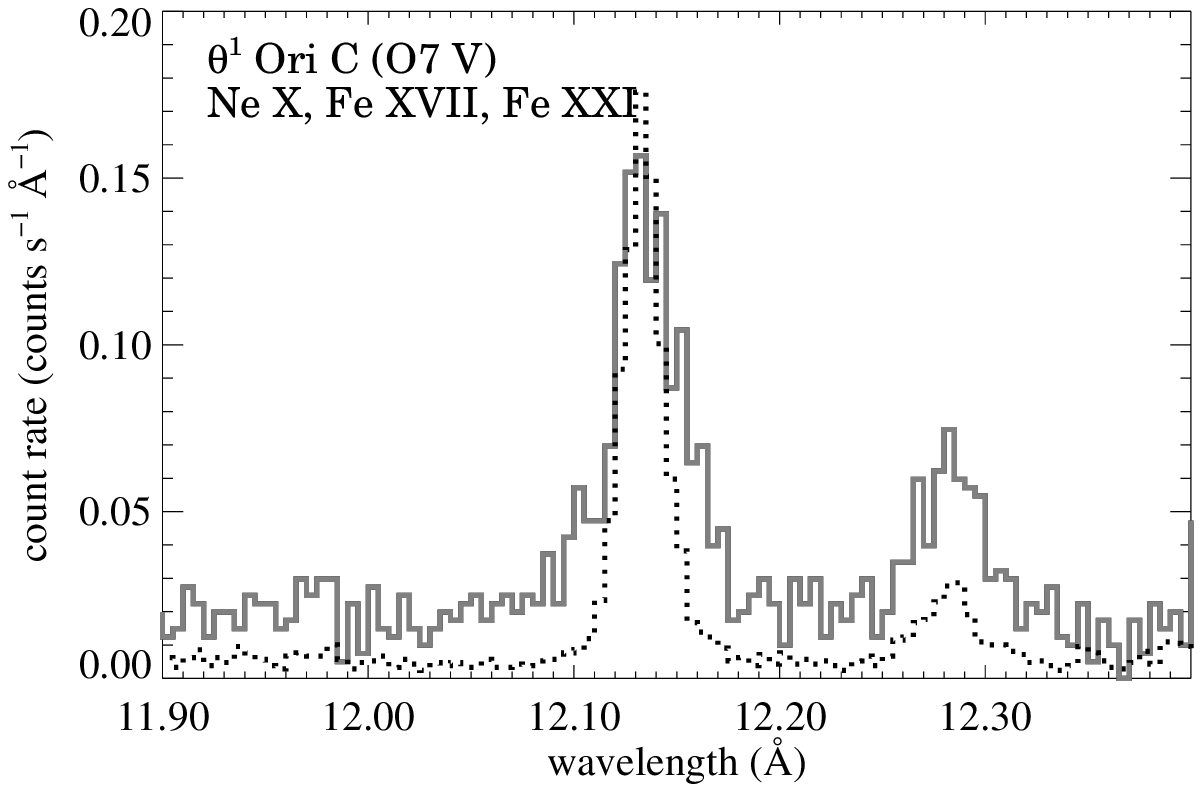}
\plottwo{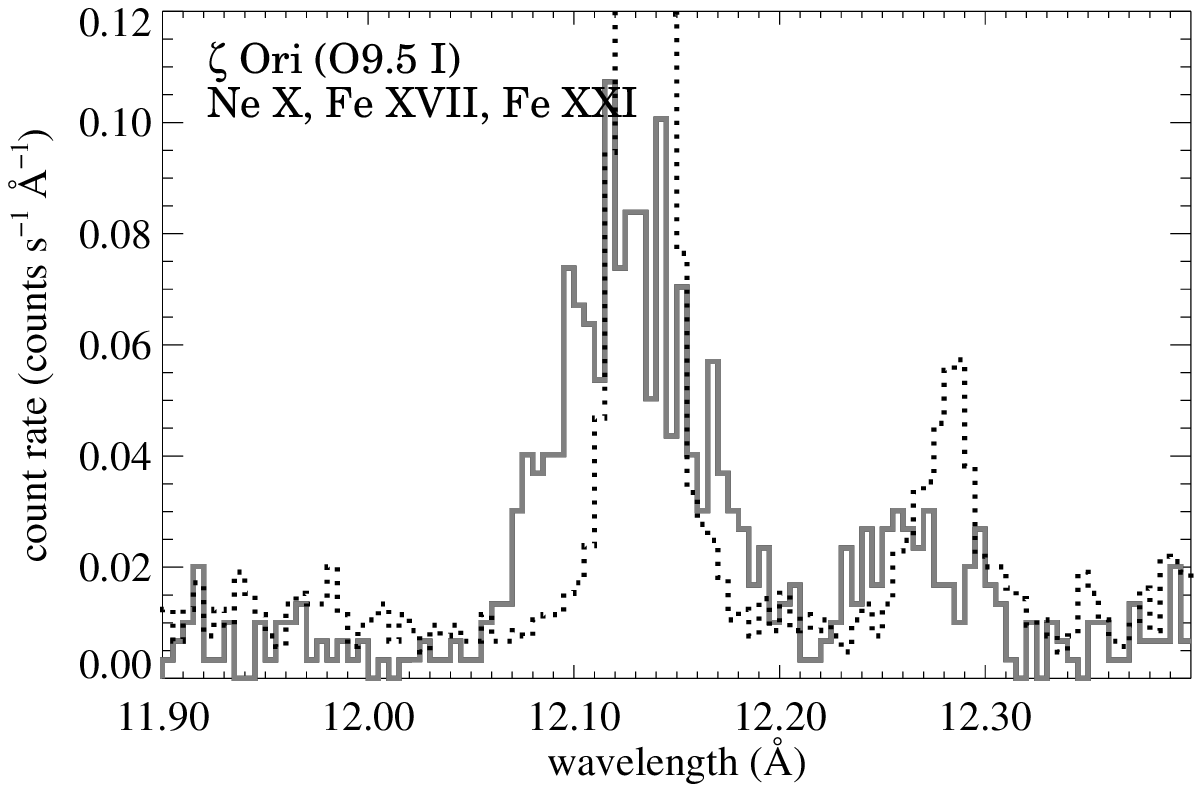}{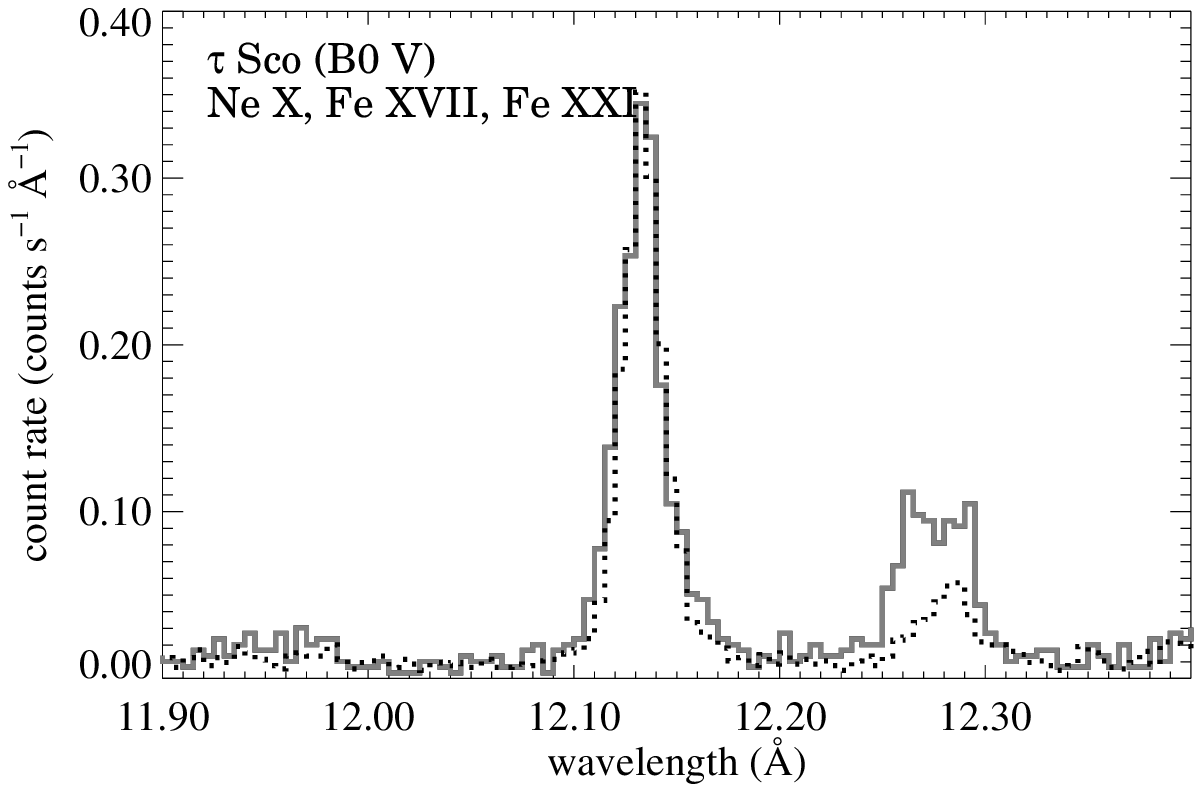}
\caption{MEG spectra (solid gray lines) of
$\zeta$~Pup (upper left panel) and $\theta^1$~Ori~C (lower panel).
The MEG spectrum of AB Dor (K0~V) is shown (dashed line) for comparison.
As expected for a luminous hot star with instability driven wind shocks,
the Ne X resonance line of $\zeta$~Pup is blueshifted, broad, and asymmetric.
The Ne X line of $\theta^1$~Ori~C (upper right panel) is symmetric
about line center and its non-thermal, non-instrumental width is
$v \approx 410{\rm~km~s}^{-1}$.  Note the strong Fe XXI line.
$\zeta$~Ori (lower left panel) shows a broad Ne X line.  The $\tau$~Sco lines
(lower right panel) are narrow, like those of the coronal source AB Dor (dashed line).
}
\end{figure}

We note that T Tauri stars have strong UV emission lines (H~I, N~V,
C~IV, Si~IV) produced by optically thin $5\times10^4-2\times10^5$~K plasma.
The continuum emission (which depopulates the $^3S_1$ state) is not
produced by an optically thin transition region.  If this thick material
is distributed over much of the photosphere, the UV continuum would be too
diluted to suppress the O~VII and Ne~IX forbidden lines.
The forbidden lines are suppressed if the X-rays are
emitted at the base of the accretion funnel directly above the optically
thick UV hotspot.

An alternate hypothesis is that X-ray emitting magnetic loops are distributed
throughout an active corona. In this case, the $f/i$ ratios are unaffected by
UV flux and $\log n_{\rm e} \approx 12.75\rm{~cm}^{-3}$,
the total particle density $n = 1.77 n_{\rm e}$, the mean temperature in
the loop is at the peak of the emission-measure distribution 
$T \approx 3.0\times10^6\rm{~K}$, and the total gas pressure
in the loop is $P = nkT = 4.12\times10^3\rm{~dynes~cm}^{-2}$.
The mean magnetic field strength required to confine such a loop is
$B^2 \geq 8\pi n k T \Rightarrow B \geq 320\rm{~G}$. Direct measurements
of magnetic field strengths and filling factors on other T Tauri star
indicate very strong 1--4~kG fields concentrated in cool
spots covering a large fraction of the photosphere (Johns-Krull, Valenti, \&
Koresko 1999; Guenther et al. 1999). Even allowing for the geometric reduction
in the magnetic intensity as loops expand into the corona, 
coronal loops on T Tauri stars can sufficiently confine high-density
coronal plasma.

In conclusion, X-ray and far-UV spectra of T Tauri
stars may provide a key test of accretion-powered X-ray activity.
If the forbidden lines are suppressed by photoexcitation, Si~XIII,
Mg~IX, Ne~IX, and O~VII $f/i$ ratios will yield
self-consistent estimates of the UV/X-ray source separation.

\section{X-ray Grating Spectra of OB stars}

The far-ultraviolet flux from OB stars is $10^5-10^6$ times brighter than
on T Tauri stars like TW Hya. 
As a result, O~VII and Ne~IX f-line emission is essentially absent.
The $^3{\rm S}_1 \rightarrow {^3{\rm P}}$ transitions of Mg~XI and Si XIII
at 1024.4~\AA\ and 901.8~\AA, respectively, are used to estimate the average
distance bewteen the photosphere and the X-ray emitting region. 
Kahn et al. (2001) and Cassinelli et al. (2001) have shown for $\zeta$~Pup (O4~If)
that the visible X-rays are emitted $1-5 R_{\star}$ from the photosphere where the
overlying wind becomes optically thin at the wavelengths of Mg~XI, Si~XIII, and S~XV.
These results indicate that X-ray emitting regions are distributed throughout
the wind as predicted by instability-driven wind shock theory. $\zeta$~Pup's X-ray
line profiles provide further confirmation of this picture.  In the upper left
panel of Figure~3, the 12.1~\AA\ Ne~X line of $\zeta$~Pup is blueshifted, broad,
and asymmetric because the optically thick wind absorbs most of the X-rays from
receding shocked material on the far side of the star.
The challenge for wind shock models is to produce strong enough shocks at the
base of the wind to explain the Si~XIII and S~XV emission.

While high-resolution spectra of $\zeta$~Pup support the distributed wind-shock
hypothesis, HETG spectra of three other hot stars are more difficult to explain.
In Fig. 3, the Ne~X line of $\theta^1$~Ori~C (O7~V) and
$\tau$~Sco (B0~V) do not show obvious line asymmetries.
In their analysis of the $\theta^1$~Ori~C spectrum, Schulz et al. (2000) find a
range of line widths ($300-800{\rm~km~s}^{-1}$) and no systematic line shifts.
Schulz et al. (2000) state that these results are consistent with
instability-driven wind shocks.  However, Owocki \& Cohen (2001) have shown
that the substantial amount of emission longward of line center is
difficult to reconcile with the expected attenuation by the wind.
Based on $f/i$ ratios, Waldron \& Cassinelli (2001) find that the
hottest plasma around $\zeta$~Ori must lie less than $1 R_{\star}$ from the
photosphere, concluding that wind shocks and some magnetic confinement of
turbulent hot plasma may be required. The $\tau$~Sco lines are narrow and may
not be produced in outflowing material many stellar radii from the
photosphere.

\begin{figure}
\plotone{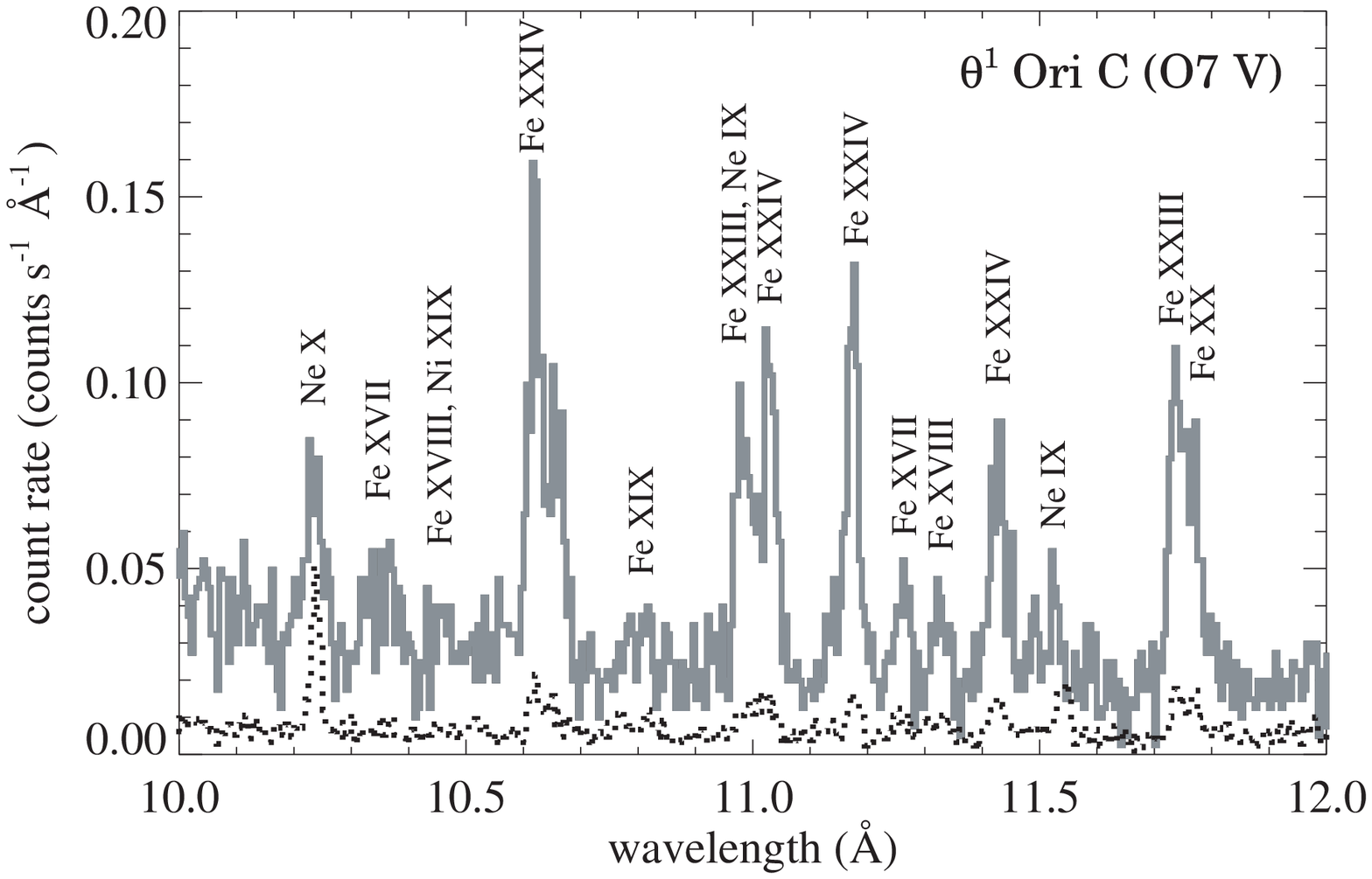}
\plotone{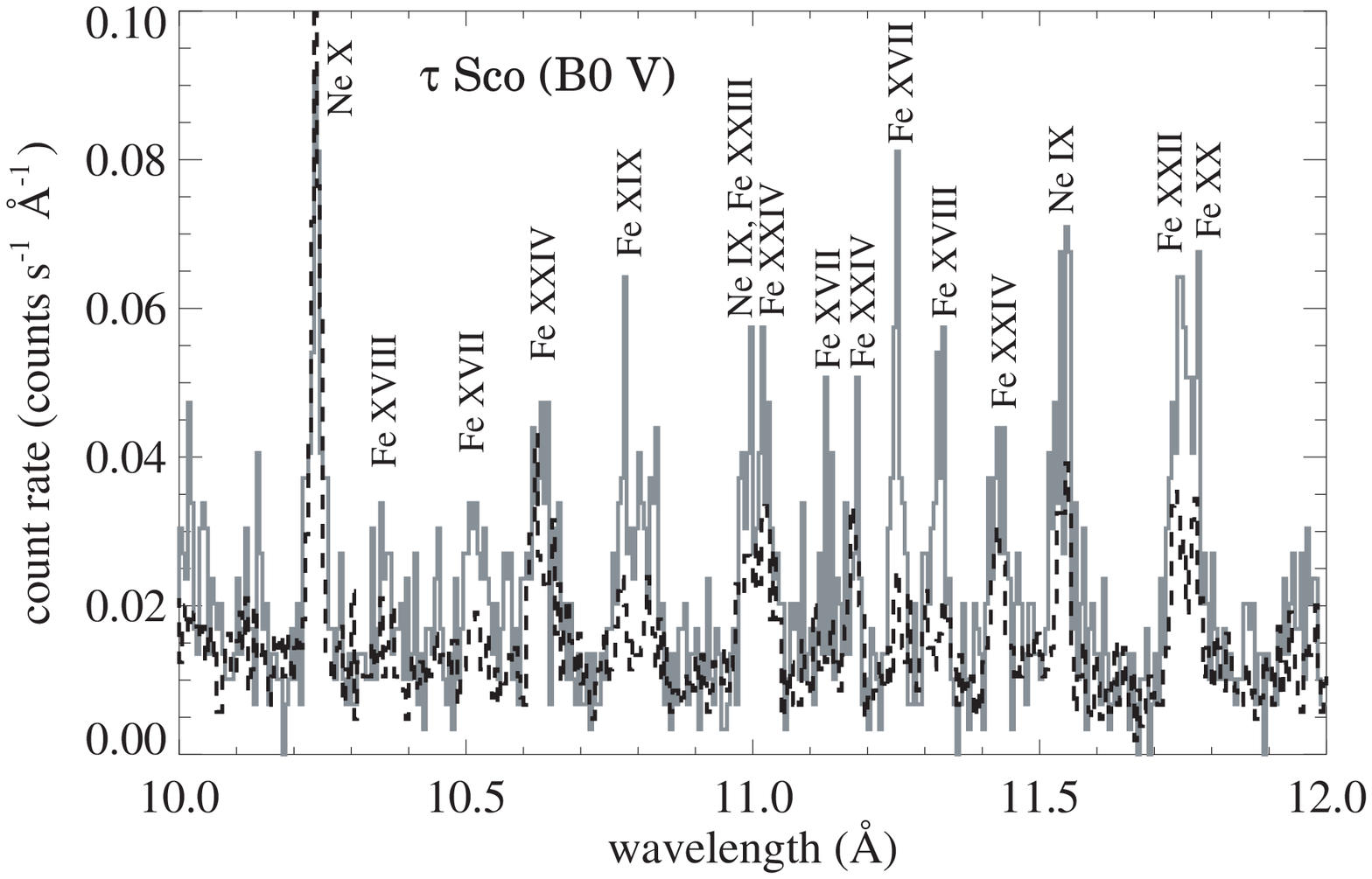}
\caption{Upper panel: the MEG 10--12 \AA\ spectrum of $\theta^1$~Ori~C
shows strong, moderately broad, symmetric Fe~XX--Fe~XXIV lines indicating
substantial emission measure at or above 16~MK.
Lower panel: the $\tau$~Sco (B0~V) spectrum shows narrow Fe~XX--Fe~XXIV
lines similar to those seen on the active corona of AB Dor (K1~V) (dashed lines).}
\end{figure}

The 10--12 \AA\ spectra of $\zeta$ Pup, $\theta^1$~Ori~C, and $\tau$~Sco also
exhibit striking differences. $\zeta$ Pup (see Figure~3 of Cassinelli et al.
2001 and Figure~5 of Kahn et al. 2001) shows broad, asymmetric,
blue-shifted lines of Ne~IX and Fe~XVII, indicating little emission measure
above 10~MK.  The 10--12 \AA\ MEG spectra of $\theta^1$~Ori~C and $\tau$~Sco
in Figure~4 show very strong, narrow, symmetric Fe~XXIII and Fe~XXIV lines
suggesting very hot ($T>16$~MK), solar-abundance plasma (Schulz et al. 2000).
To date, instability-driven wind shock models
cannot produce such high-temperature shocks, especially
not on OB stars like $\theta^1$~Ori~C and $\tau$~Sco with relatively
low mass-loss rates and terminal wind speeds.

\section{Magnetically Confined Wind Shocks}

The leading theoretical picture for X-ray production in magnetized hot stars is
the magnetically confined wind shock (MCWS) model proposed by Babel \&
Montmerle (1997a) and applied by the same authors to $\theta^1$~Ori~C (1997b).
In this model, an ionized, radiation-driven wind is confined to flow along
a magnetic dipole field and is thus channeled, at low latitudes, into the
magnetic equatorial plane.  In the
magnetic equator, the streams from the two hemispheres collide and are
confined by the dipole field, producing a standing shock and X-ray emission
above and below a disk-like feature above the magnetic equator.

This model succeeds in producing relatively hard X-rays (due to the nearly
head-on collision of two fast streams), high levels of emission (due
to the confinement at relatively high densities of the shock-heated plasma),
and rotational modulation of the X-rays, like that observed in
$\theta^1$~Ori~C, if the dipole field is tilted with respect to the rotational
axis.  This oblique magnetic rotator model has been used to explain many of
$\theta^1$~Ori~C's curious behaviors: periodic H$\alpha$ and
He~II emission and periodic C~IV and Si~IV P-Cygni profiles (Stahl et al. 1993;
Walborn \& Nichols 1994; Stahl et al. 1996; Reiners et al. 2000).
Based on circular spectro-polarimetric monitoring of its Balmer lines,
Jean-Fran\c{c}ois Donati and collaborators have recently measured a polar
magnetic field strength $B\approx1.1$~kG (2001, private communication)
at phase 0.0.  This implies that X-ray, H$\alpha$ and
He~II emission is greatest when looking down on the magnetic pole (phase 0.0).
Emission is weakest when looking at the magnetic equator (phase 0.5),
presumably because hot plasma just above the magnetic
equator is eclipsed by the photosphere.  We note that
Stahl et al. and Babel \& Montmerle proposed the opposite magnetic geometry. 

\begin{figure}
\centerline{\includegraphics[width=0.6\textwidth]{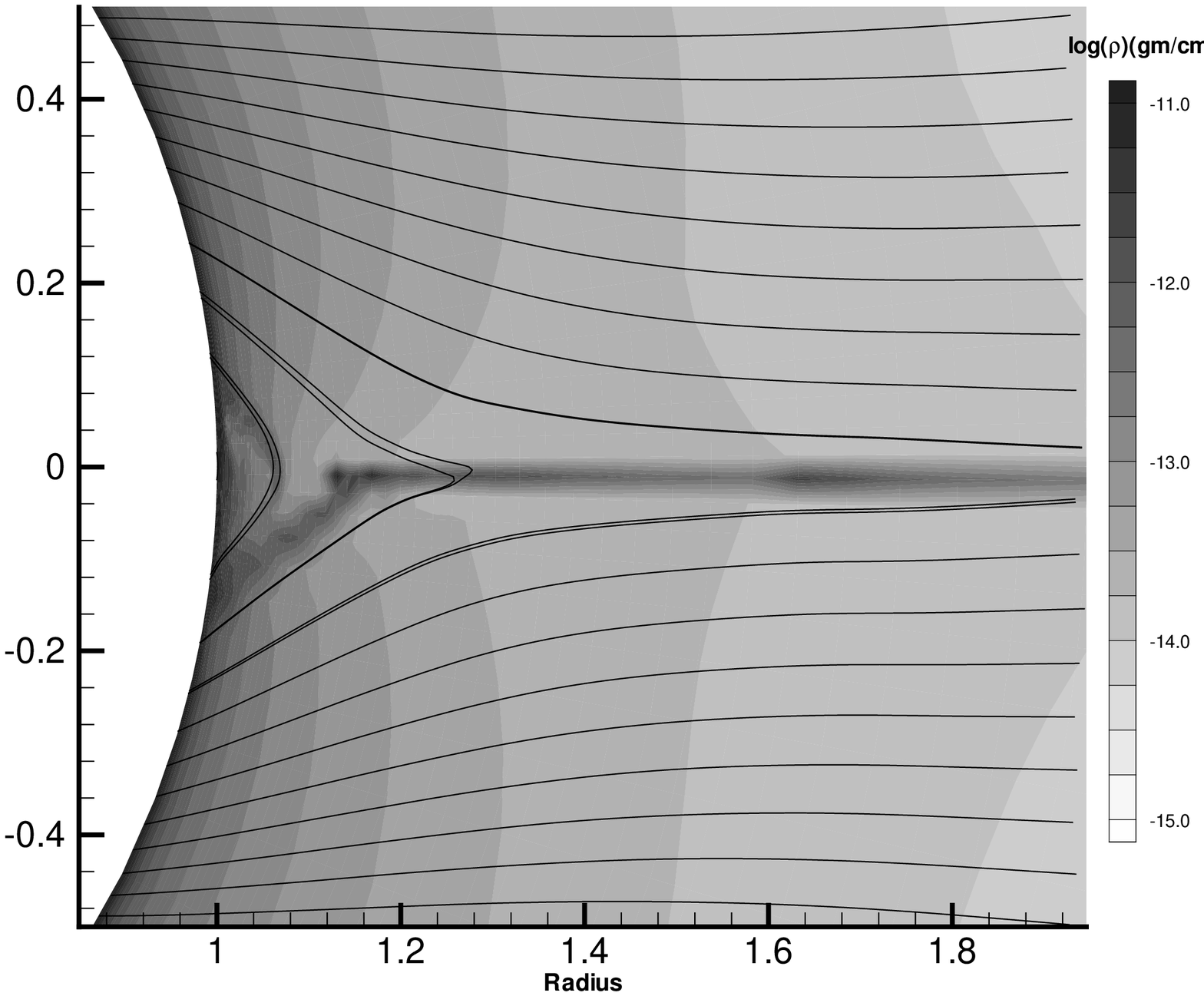}}
\centerline{\includegraphics[width=0.6\textwidth]{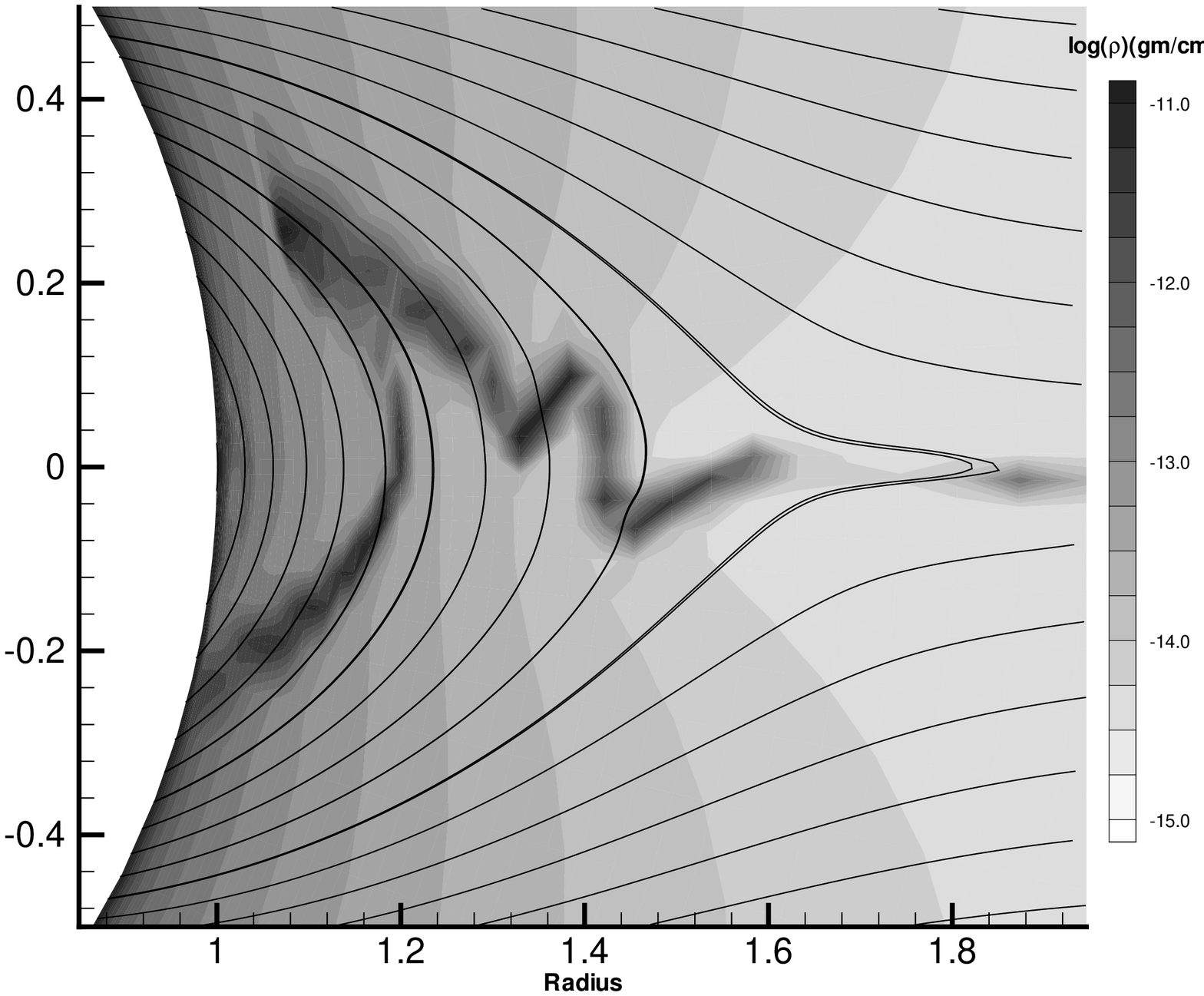}}
\caption{2-D MHD simulations of an initially dipolar magnetic field distorted
by radial mass loss on $\zeta$~Pup. Density (upper panels) and
non-radial velocity contours (lower panels) at the end of two realizations:
moderate magnetic confinement ($\eta = \sqrt{10}$, $t = 450$~ks) on the left
and high confinement ($\eta = 10$, $t = 295$~ks) on the right.
With sufficient confinement, material is accelerated to approximately
$\pm 1000{\rm~km~s}^{-1}$ creating shocks in dense, unstable regions near
the magnetic equator only $0.1-1R_{\star}$ above the photosphere.
The recent measurement of a 1.1-kG polar field on $\theta^1$~Ori~C implies
$\eta \approx 13$ as on the right.  The resulting dense, high-speed
shocks may explain the broad, symmetric, high-temperature lines seen
on $\theta^1$~Ori~C.
}
\end{figure}

\begin{figure}
\centerline{\includegraphics[width=0.5\textwidth]{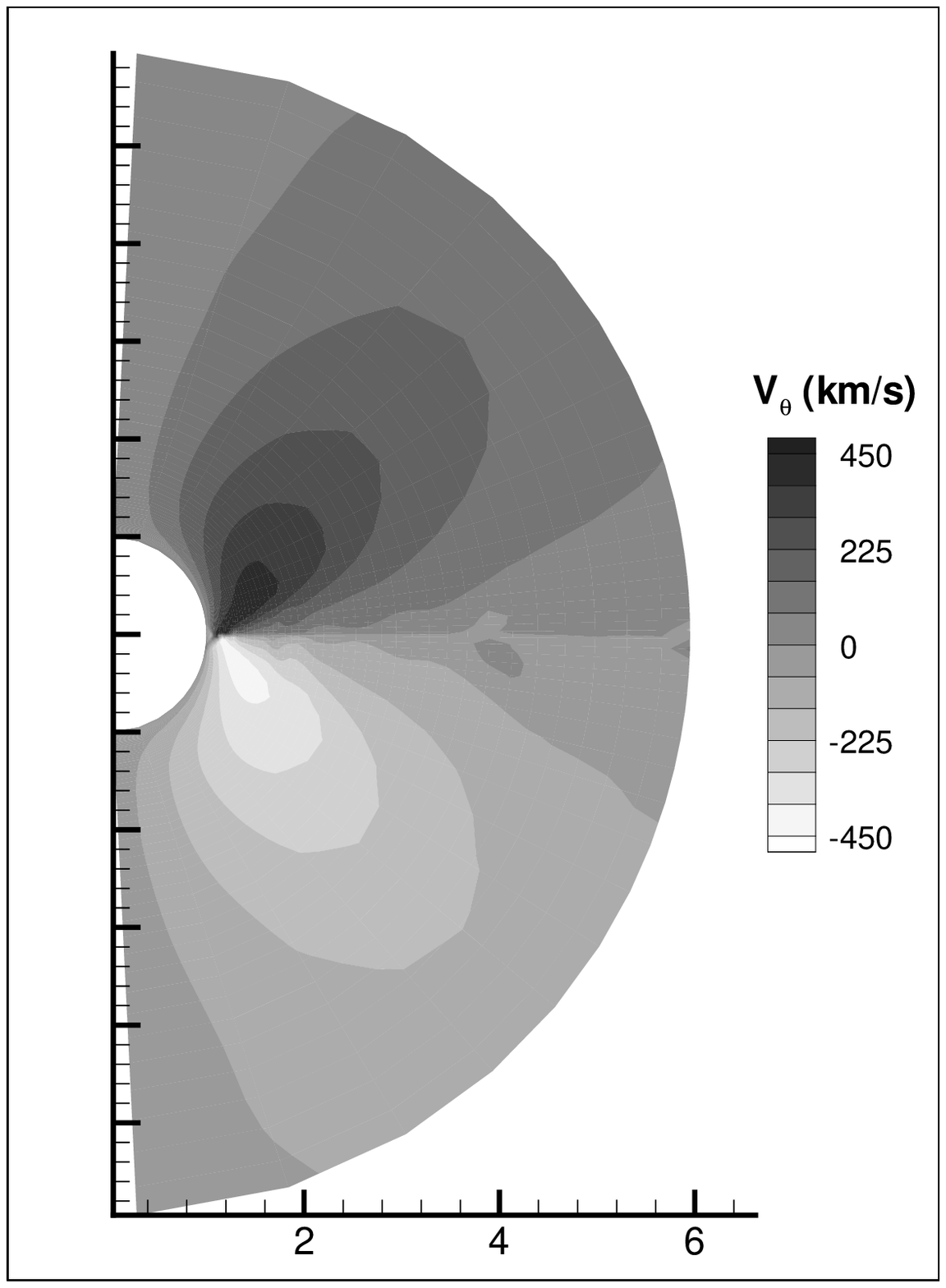}}
\centerline{\includegraphics[width=0.5\textwidth]{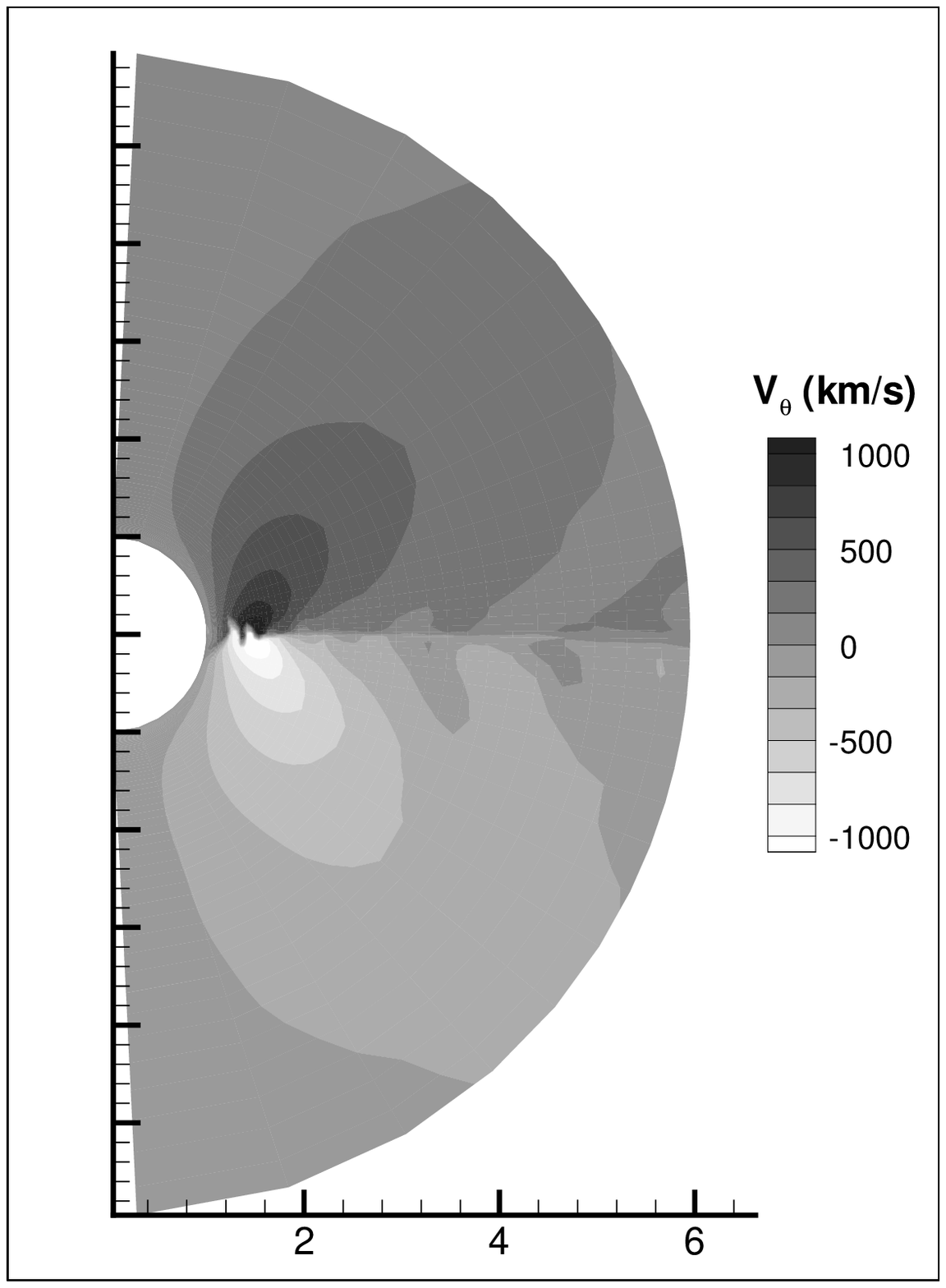}}
\end{figure}

The model of Babel \& Montmerle, which assumes a rigid magnetic geometry,
has been extended to allow the magnetic field morphology to be modified by the wind
flow in a self-consistent manner (Ud-Doula \& Owocki 2001, in preparation).
Assuming an initially dipolar field and spherical mass-loss, the 2-D MHD code
simulates the time evolution of density, temperature, and velocity throughout the wind.
A given set of field strength and wind parameters defines a confinement parameter, 
\[ \eta = \frac{B_0^2 R_\star^2}{\dot{M} v_{\infty}}, \]
where $B_0$ is the surface equatorial field strength (approximately half
the polar field strength).
$\eta \gg 1$ implies the wind is confined by the field and $\eta \ll 1$
implies the field is completely stretched out by the wind.

In Figure~5, we show a pair of $\zeta$~Pup simulations for
$\eta = \sqrt{10}$ and $\eta = 10$.  For $\theta^1$~Ori~C, $B_0\approx 550~G$
and $\eta \approx 13$, implying high-speed shocks, $T \ga 10^7$~K,
and $n_{\rm e} \ga 10^{10}{\rm~cm}^{-2}$.  As of this writing, we do
not know if the wind continuum optical depth is low enough to view
the entire volume of X-ray emitting plasma.  Nonetheless, the 2-D
simulations suggest that a moderately strong dipolar field can provide
significant confinement and shock heating.
To realistically test the MCWS idea, synthetic X-ray spectra based
on 3-D MHD simulations of an oblique magnetic rotator are needed for a range
of mass-loss rates, magnetic field geometries, and viewing angles.

\section{Discussion}

The velocity profiles, $f/i$ ratios, and high plasma temperatures
derived from {\it Chandra} spectra of OB stars suggest that magnetic fields may
play an important role in the production of X-rays on young, massive stars.
Time-resolved, high-resolution X-ray spectra of very young stars may help
uncover the geometry of the underlying field.
It is interesting to note that S1~Oph, $\theta^1$~Ori~C, and $\tau$~Sco,
which show the most striking signs of magnetic activity, are all young 
and associated with star-forming regions: 
$\rho$~Oph ($\la 1$~Myr, Andr\'e et al. 1988),
Orion Nebula ($\la 1$~Myr, Hillenbrand 1997),
and Sco-Cen ($\la 10$~Myr, Kilian 1994).
Although a different kind of magnetic activity
produces very hot, energetic X-ray flares seen on various classes of low-mass
YSOs, both high-mass and low-mass stars undergo a brief episode of rapid
accretion and mass-loss regulated by large-scale fields and disks. 
The evolution of the field and disk affects angular momentum evolution,
planet formation, and the high-energy radiation environment of
young planetary atmospheres.

\end{document}